\documentstyle[11pt,appb2]{article}

\newcommand{\Fh}[2]{\,{}_#1F_#2}
\newcommand{\Fs}[3]{\!\!\left[\begin{array}{c}#1\,;\\#2\,;\end{array}#3\right]}

\newcommand{\Fup}[2]{\Fs{#1}{#2}{\frac{q^2}{m^2}}}

\begin{document}


\title{REDUCTION OF FEYNMAN GRAPH AMPLITUDES\\
       TO A MINIMAL SET OF BASIC INTEGRALS%
\thanks{Presented at the DESY Zeuthen Workshop on Elementary
    Particle Theory ``Loops and Legs in Gauge Theories'', Rheinsberg,
    Germany, April 19-24, 1998.}}

\vspace{0.5cm}

\author{
 O.V.Tarasov\thanks{
 On leave of absence from JINR, 141980 Dubna (Moscow Region), Russian
 Federation.
Address after October 1st, 1998: ITP, University of Bern, Switzerland}
\address{
DESY Zeuthen, Platanenallee 6, D--15738 Zeuthen, Germany\\
E-mail: {\tt tarasov@ifh.de}
}}

\vspace{0.5cm}

\maketitle
\begin{abstract}
An algorithm for the reduction of massive Feynman integrals 
with any number of loops and external momenta  to a minimal
 set of basic integrals is proposed. The method is based on the 
new algorithm for evaluating tensor integrals,  representation
of generalized recurrence relations \cite{OVT1} for a given
kind of integrals as a linear system of PDEs and  the  reduction
of this system to a standard form using algorithms proposed
in \cite{Reid1}, \cite{Reid2}.
Basic integrals reveal as parametric derivatives of the
system in the standard form and the number of basic integrals
in the minimal set is determined by the dimension of the solution
space of the system of PDEs.
\end{abstract}

\vspace{0.5cm}

\section{Introduction}

Mass effects play an important role in confronting experimental data
obtained at the high-energy colliders, like LEP and SLC, with
theoretical predictions. Precise determinations of many physical
parameters in the Standard  Model (SM) require the evaluation of
mass dependent radiative corrections.
 In the SM, due to the complicated structure of
integrals and the large number of diagrams (typically thousands),
no complete calculation even of two-loop self-energy has been carried
out. Many different species of particles with different masses have
to be taken into account and this makes the evaluation of multi-loop
integrals a rather difficult problem. Existing  numerical methods
for evaluating Feynman diagrams cannot guarantee the required accuracy
for the sum of  thousands of  diagrams and therefore  the
development of analytical or semi-analytical methods are of
great importance.

In calculating Feynman diagrams mainly  three  difficulties arise:
tensor decomposition of integrals, reduction of scalar integrals to
several basic  integrals and the evaluation of these basic integrals.

For the reduction of scalar integrals to a minimal set of integrals,
recurrence relations are the most suitable tool. Recurrence 
relations originating from the method of 
integration by parts \cite{CT} were very helpful  in various
sophisticated calculations.
Up to now all attempts to extend the standard method of integration
by parts  to multi-loop diagrams with arbitrary external
momenta in the case when all the masses are different were unsuccessful.
The first successful algorithm for the reduction of two-loop massive
propagator type integrals to a minimal set of integrals
\cite{OVT2} was based on generalized recurrence relations proposed in
ref.  \cite{OVT1}.

Using the method described in \cite{OVT1}, it is easy to write 
down an enormous number of  generalized recurrence relations for any
kind of integrals. Not all recurrence relations are independent
and therefore a prescription for determining an optimal set
of recurrence relations is highly desirable. Another
closely related and important  problem  is the determination of
the minimal set of basic integrals.

 In the present paper we propose an approach which in principle
completely  solves the problems mentioned before. 
We formulate an algorithm for transforming tensor integrals into a
combination of scalar ones.
We show how to reduce systems of generalized recurrence relations
to a linear system of PDEs.
The algebraic analysis of linear systems of
PDEs is an intensely developing branch of contemporary mathematics
and many existing algorithms and methods in this field can be used 
for our purposes. The reduction of  linear systems
of PDEs to an involutive or standard form or to a
differential Gr\"obner bases \cite{Volodya}  solves the problem
of finding the optimal set of recurrence relations and
a minimal set of basic integrals. Involutive system
of equations can be used for the analytic or numeric
evaluation of basic integrals or finding different kind
of asymptotic expansion. 

The paper is organized as follows.
In Sec. 2 a method for the representation of tensor integrals and
integrals with irreducible numerators in terms of scalar ones with
shifted space-time dimension $d$ is shortly described.
In Sec. 3 we describe the method of generalized recurrence relations.
In Sec. 4 we show how to transform the system of generalized
recurrence relations into a linear system of PDEs.
In Sec. 5 we describe the main features of the Standard Form
algorithm by G. Reid and explain the correspondence 
between the standard form of the linear system of PDEs
and our recurrence relations.

\section{Evaluation of multi-loop tensor integrals }


For the evaluation of $L$ loop Feynman graph amplitudes one needs to calculate the
integrals
\begin{equation}
G^{(d)}(\{q_iq_k\},\{m_s^2\})=
\prod_{i=1}^{L} \int d^dk_i \prod^{N}_{j=1}
P^{\nu_j}_{\overline{k}_j,m_j}\prod_{r=1}^{n_1}
k_{1\mu_r} \ldots \prod_{s=1}^{n_L}
k_{L\lambda_s} ,
\label{arbdia}
\end{equation}
where
\begin{equation}
P^j_{k,m}=\frac{1}{(k^2-m^2+i \epsilon)^j},~~~~~~~~~~~
\overline{k}_j^{\mu}=
 \sum^L_{n=1} \omega_{jn}k^{\mu}_n+ \sum _{m=1}^E
 \eta_{jm}q_m^{\mu}
\end{equation}
and $q_m$ are external momenta,  $N$ is the number of lines,
$E$ is the number of external legs, $\omega$ and $\eta$ are matrices
of incidences of the graph with the matrix elements being $\pm1$ or $0$.

The traditional way to evaluate tensor integrals consists
of the following steps:
\begin{itemize}
\item write for the integral the most general tensor
as a polynomial in terms of monomials in external momenta
and the metric tensor $g_{\mu \nu} $
\item multiply this tensor by appropriate tensor
monomials and obtain a linear system of equations
\item solve the system of algebraic equations
\end{itemize}
One-loop tensor integrals can always be reduced to a combination
of scalar integrals without scalar products in the numerator.
Evaluating multi-loop integrals one encounters so-called
irreducible numerators, i.e. scalar products which cannot be
expressed in terms of scalar factors in the denominator.
The solution of this problem for multi-loop integrals as well as
for the evaluation of multi-loop tensor integrals was
proposed in \cite{OVT1}. The main idea of this method consists in 
the representation of tensor integrals in terms of scalar ones 
with shifted space-time dimension $d$. For tensor integrals with 
any number of loops, internal lines and external momenta, 
the following formula was proposed:
\begin{eqnarray}
\prod_{i=1}^{L} \int d^dk_i \prod^{N}_{j=1}
P^{\nu_j}_{\overline{k}_j,m_j}\prod_{r=1}^{n_1}
k_{1\mu_r} \ldots \prod_{s=1}^{n_L}
k_{L\lambda_s} &=& \nonumber \\
T_{\mu_1 \ldots \lambda_{n_L}}(\{q_i\},\{ \partial_j  \}, {\bf d^+})
\prod_{i=1}^{L} \int d^dk_i \prod^{N}_{j=1}
P^{\nu_j}_{\overline{k}_j,m_j},
\label{tensint}
\end{eqnarray}
where $T_{\mu_1 \ldots \lambda_{n_L}}$ is a polynomial type
tensor operator and
\begin{equation}
\partial_j \equiv \frac{\partial}{\partial m_j^2},
~~~~~~~~{\bf d^+} G^{(d)}=G^{(d+2)}.
\end{equation}

The main ingredients of the derivation of the operator $T$
are independent auxiliary
vectors $a_i (i=1,\ldots, L)$ and the  use of the  $\alpha$- parametric
representation. The tensor structure of the integrand on
the left-hand side of (\ref{tensint})  can be written as
\begin{equation}
k_{1\mu_1} \ldots  k_{L\lambda_{n_L}}=
\left.\frac{1}{i^{n_1+...+n_L}}\frac{\partial}{\partial a_{1\mu_1}}
  \ldots \frac{\partial}{\partial a_{L\lambda_{n_L}}}
 \exp \left[i (a_1k_1\!+\! \ldots+\! a_Lk_L)\right]
\right|_{ a_i=0 }.
\label{avectors}
\end{equation}
To convert the integral
\begin{equation}
G^{(d)}=\prod_{i=1}^{L} \int d^dk_i \prod^{N}_{j=1}
P^{\nu_j}_{\overline{k}_j,m_j} \exp{i[a_1k_1 + \ldots +a_Lk_L]},
\end{equation}
into the $\alpha$-representation, we first transform all 
propagators into a parametric form
\begin{equation}
\frac{1}{(k^2-m^2+i\epsilon)^{\nu}}
 = \frac{i^{-\nu}}{ \Gamma(\nu)}\int_0^{\infty}
 d\alpha ~\alpha^{\nu-1} \exp\left[i\alpha(k^2-m^2+i\epsilon)\right].
\end{equation}
Using the $d$-dimensional Gaussian integration  formula
\begin{equation}
\int d^dk \exp \left[i(A k^2+ 2(pk))\right] =i
 \left( \frac{\pi}{i A} \right)^{\frac{d}{2}}
 \exp \left[ -\frac{ip^2}{A} \right] ,
\end{equation}
we  evaluate the integrals over loop momenta and obtain
\begin{eqnarray}
G^{(d)}&=&i^L \left(\frac{\pi}{i} \right)^{Ld/2} \prod_{j=1}^{N}
\frac{i^{-\nu_j}}{\Gamma(\nu_j)} \int_{0}^{\infty}
\ldots \int_{0}^{\infty}
\frac{d \alpha_j \alpha_j^{\nu_j-1}}{[D(\alpha)]^{d/2}}
\nonumber \\
&& \nonumber \\
&&\times \exp{i\left[ \frac{Q(\{ q_j \}, \{\alpha_j \}, \{a_j \})}
{D(\alpha)} -\sum_{i=1}^N \alpha_i(m_i^2-i\varepsilon) \right]},
\label{alfrep}
\end{eqnarray}
where $D$ and $Q$ are polynomials in $\alpha,~a$ and $q_iq_j$.
Differentiating (\ref{alfrep}) with respect to $a_j$ we get
by simple identification the operator $T$:
\begin{eqnarray}
T(\{q\},\{\partial \},{\bf d^+})
&=&\frac{e^{-iQ(\{ \{q\}, \alpha, \{0\} ) \rho}
 }{i^{n_1+ \ldots +n_N}}
\!\prod_{r=1}^{n_1} \frac{\partial}{ \partial a_{1\mu_r}} \!\!\
\ldots \nonumber \\ 
&& \prod_{s=1}^{n_L} \left. \frac{\partial}{ \partial a_{L\lambda_s}}
e^{i[ Q(\{ q_j\},\alpha,\{a_j\})] \rho }
\right|_{ a_j=0~~~~~~\atop {\alpha_j=i \partial_j \atop
  \rho=(\frac{i}{\pi})^L{\bf d^+}} }.
\label{Ttensor}
\end{eqnarray}

For any particular tensor integral the operator $T$ can be 
constructed by using computer algebra languages.
 To generate $T$ for 2-3 loop tensor integrals of rank
3-4 with the help of a short  package written
in FORM, several minutes on a PC Pentium 90 were needed.
To our opinion the evaluation of tensor integrals by the proposed
method has several advantages in comparison with the traditional
method. Firstly, in order to obtain the tensor decomposition
no contractions with external momenta and the metric tensor
and no solution of a linear system of equations
are needed. Secondly, it is easy to select the scalar coefficient
of the particular tensor structure.
Thirdly, a representation in terms of integrals with shifted
$d$ (as we have seen on many one- and two-loop examples) 
 is very compact and may be useful for numerical calculations.

To obtain the final result, scalar integrals with different
indices and different shifts in $d$ are to be evaluated.
This problem  can be solved by using the method of generalized
recurrence relations as proposed in \cite{OVT1}.


\section{Generalized recurrence relations}


In order to obtain recurrence relations for integrals
we use  identities like:
\begin{equation}
\prod_{i=1}^{L} \int d^dk_i
\frac{\partial }{\partial k_{r\mu}}
\left\{ R_{\{ \mu\} }\left( \{ k \}, \{ q \} \right)
\prod^{N}_{j=1}
P^{\nu_j}_{\overline{k}_j,m_j} \right\} \equiv 0,
\label{ibpm}
\end{equation}
where $R$ is an arbitrary tensor polynomial.
After performing the differentiation, two different 
representations for scalar products can be used:
\begin{eqnarray}
&& \nonumber \\
&&{\rm a}) ~~~k_iq_j=\frac12 (k_i^2+q_j^2-(k_i-q_j)^2), \nonumber
\\
&& \nonumber \\
&& {\rm b} )~~~
 q_{j\mu} \int k_{i \mu} d^dk_i  \prod^{N}_{j=1}
P^{\nu_j}_{\overline{k}_j,m_j} =
q_{j \mu} T_{\mu}( \{ q \}, \{ \partial \}, {\bf d^+} )
\int d^dk_i \prod^{N}_{j=1} P^{\nu_j}_{\overline{k}_j,m_j}.
\end{eqnarray}

By using all possible combinations of these
representations, we  produce many
relations connecting integrals with changed exponents of scalar
propagators  and changed values of the  space-time dimension.
For the same scalar product both a) and b) can be used.
Combining the different relations,  one can try to find the most
 optimal set of relations for the reduction of the concrete class
of integrals to the minimal set of basic integrals.

In fact, in the traditional method of integration by parts
only representation a) for scalar products is used.
Our derivation is more general and it includes the integration
by parts method \cite{CT} as a particular case.
Recurrence relations connecting integrals with shifted $d$
cannot be obtained from the traditional
method of integration by parts.

To illustrate the difference, we consider
the one-loop propagator type integral with massive particles:

\begin{equation}
I^{(d)}_{\nu_1 \nu_2} = \int  \frac{d^d k_1}
{[i\pi^{d/2}]} P_{k_1,0}^{\nu_1} P_{k_1-q,m}^{\nu_2}.
\label{i1ab}
\end{equation}
From the traditional method of integration by parts
two relations can be derived:
\begin{eqnarray}
2\nu_2 m^2 I^{(d)}_{\nu_1~\nu_2\!+\!1}
+\nu_1 I^{(d)}_{\nu_1-1~ \nu_2+1}
+\nu_1 (m^2\!-\!q^2) I^{(d)}_{\nu_1+1~ \nu_2}
\nonumber \\
-(d-2\nu_2-\nu_1) I^{(d)}_{\nu_1 \nu_2}&=&0,
\nonumber \\
\nu_1 I^{(d)}_{\nu_1\!+\!1~ \nu_2-1}\!-\!\nu_2  I^{(d)}_{\nu_1
-1~\nu_2+1}\!
+\!\nu_1(m^2\!-\!q^2) I^{(d)}_{\nu_1+1~\nu_2}
\!\nonumber \\
\!+ \!\nu_2 (\!m^2\!+\!q^2) I^{(d)}_{\nu_1~ \nu_2+1}
+(\nu_2\!-\!\nu_1) I^{(d)}_{\nu_1 \nu_2}&=&0.
\nonumber \\
\label{CTferm}
\end{eqnarray}
The integral $I^{(d)}_{\nu_1 \nu_2}(q^2,0,m^2)$
is proportional to the
Gauss hypergeometric  function \cite{BoDa}:
\begin{eqnarray}
I^{(d)}_{\nu_1 \nu_2}(q^2,0,m^2)&=&(-1)^{\nu_1+\nu_2}
 \frac{\Gamma(\nu_1+\nu_2-\frac{d}{2}) \Gamma(\frac{d}{2}-\nu_1) }
 {(m^2)^{\nu_1+\nu_2-\frac{d}{2}} \Gamma(\frac{d}{2}) \Gamma(\nu_2)}
\nonumber \\
&& \times \Fh21\Fup{\nu_1,\nu_1+\nu_2-\frac{d}{2}}{ \frac{d}{2} }.
\label{BoDa}
\end{eqnarray}
As is well known there are fifteen relations of Gauss between
contiguous functions $_2F_1$. Substituting ($\ref{BoDa}$)
into  ($\ref{CTferm}$) one can find
correspondence between the recurrence relations (\ref{CTferm})
and only six  relations of  Gauss. The reason is obvious - in 
the relations  (\ref{CTferm}) the third parameter
of  $_2F_1$ in ($\ref{BoDa}$) does not change and therefore not 
all corresponding relations for contiguous functions can
be reproduced. If we include into consideration also the
generalized recurrence relations, we cover all fifteen relations.

Usually the number of generalized recurrence relations for
a given kind of integrals is quite substantial. 
For their effective  applications 
we must find answers to the following questions:
\begin{itemize}
\item How can one  use the generalized recurrence relations?
\item How to find the minimal set of recurrence relations?
\item How to find the minimal set of basic integrals
or how to determine the number of basic integrals?
\end{itemize}
We  propose to solve these problems by
transforming the system of generalized recurrence relations
into a (in general overdetermined) system of linear
partial differential equations (PDE) and we apply algebraic
methods being developed by mathematicians for their analysis.
Several algorithms and different computer algebra packages for
the analysis of linear systems of PDE  exist.
In our opinion the most adequate will be  the algorithm
``Standard form'' by G.~Reid \cite{Reid1,Reid2,Reid3}.

\section{Transformation of generalized recurrence relations 
into a system of linear PDEs}

The system of generalized recurrence relations can be
transformed into a linear system of PDEs by several methods.
Two procedures turn out to be most convenient.

 In the first approach recurrence relations for integrals with
arbitrary powers of propagator indices $\nu_j$ are considered.
If a term with $\nu_i-1$ occurs in the recurrence relation
then one should make the substitution
$$
\nu_i \rightarrow \nu_i+1.
$$
Thus we will obtain equivalent  recurrence relations 
connecting integrals with positive shifts of indices. 
Propagators with positive 'shifts' of indices must be
represented as:
\begin{equation}
\frac{1}{[(k_i-q_s)^2-m_l^2]^{\nu_l+r}}=
\frac{\Gamma(\nu_l)}{\Gamma(\nu_l+r)}
\frac{\partial^r}{(\partial m^{2}_l)^r}
\frac{1}{[(k_i-q_s)^2-m_l^2]^{\nu_l}}.
\end{equation}
Integrals $I^{d}$ with different shifts of the parameter of the
space-time dimension $d$ must be considered as different functions, i.e:
\begin{equation}
I^{d}=V^1, ~~~
I^{d+2}=V^2, ~~~
I^{d+4}=V^3,
 \ldots
\end{equation}
These substitutions allow one to transform
the system of recurrence relations into a linear system of
PDEs for the vector function $V \equiv \{V^1, V^2, \ldots \}$.
Such a transformation of recurrence relations into a linear
system of PDEs is convenient for the analysis of integrals
with arbitrary indices. Unfortunatly,  first approach leads
to essential technical difficulties.

In the second approach one considers a system of recurrence relations
for integrals with particular integer values of $\nu_i \geq 0$.
First, one should derive recurrence relations for the
given integral $I^{(d)}_{\nu_1 \ldots \nu_N}$ with arbitrary
$\nu$'s. From relations obtained, one needs to
derive recurrence relations for the particular sets
of integer values of $\nu_i \geq 0$ by increasing the sum of indices
$S=N, N+1, N+2, ...$ ($S=\sum_{i=1}^N \nu_i$).
In this case the system of linear PDEs will include integrals
corresponding to graphs with a different  number of lines. These
integrals are to be considered as different functions.
As in the previous case scalar propagators with $\nu_i>1$
must be represented as
\begin{equation}
\frac{1}{[(k_i-q_s)^2-m_l^2]^{r+1}}=
\frac{1}{\Gamma(r+1)}
\frac{\partial^r}{(\partial m^{2}_l)^r}
\frac{1}{[(k_i-q_s)^2-m_l^2]},
\end{equation}
and integrals with shifts in $d$ must be considered as different
functions.

As was mentioned before the system of linear PDEs for a given
integral will include the original as well as simpler integrals 
obtained by contracting lines in the original one.
For integrals with contracted lines one should also write
the system of equations. For all these integrals
one can consider different sets of $\nu$'s  which can be
classified according to the value of $S$.

The analysis by means  of ``Standard Form''
  should be started with the system
for the simplest non-zero integrals obtained from the original
integral by contracting  as many lines as possible.
Applying Standard Form algorithms the minimal set of
recurrence relations (leading derivatives) and minimal set
of basic integrals (parametric derivatives) for the simplest
integral will be found. This information will be used to
investigate more complicated integrals having more lines.
Using the terminology of \cite{Reid3} integrals with contracted
lines can be treated as classification functions.

The second approach to our opinion is more suitable for
considering integrals occurring in calculating Feynman diagrams.
In the first approach with arbitrary non-integer
indices one can expect more integrals in the basic  set.

\section{Algorithm Standard Form}

As we already mentioned the algorithm Standard Form (SF) was
proposed by G.~ Reid in \cite{Reid1,Reid2} and
was implemented as {\tt Maple} package in \cite{Reid3}.
In this section  a short review of the SF algorithm will
be given. More detailed information can be found in
\cite{Reid1,Reid2,Reid3}.

In order to compare different differential equations and different
terms in a differential equation one should introduce
total ordering which will be denoted as $>_s$. The concept of
total ordering is very important.

Let us introduce several shorthands:
\begin{eqnarray}
&&a=(a_1, \ldots , a_m), \nonumber \\
&&D_aV^p=\frac{\partial^{a_1+\ldots +a_m}}
              {\partial x_1^{a_1} \ldots \partial x_m^{a_m}}V^p,
   \nonumber \\
&&{\rm ord}(a)=a_1+ \ldots a_m \geq 0, ~~~~~~~
{\rm order ~~of ~~the~ ~derivative}
\end{eqnarray}

The most complete list of different orderings the reader
can find in \cite{RR}. By default in the SF package
Tresse ordering \cite{Tresse} is used. 
 We say that $D_aV^p>_s D_bV^q$ if
\begin{tabbing}
\quad\=     \quad\= (i) \quad\=  $ ord(a) > ord(b)$ \\
or \> $~~~$ \> (ii) \> $ord(a)=ord(b),~~ {\rm and}~~ p<q,$   \\
or \> $~~~$ \> (iii)~ $ord(a)=ord(b),~~ $p=q$ ~~{\rm and}~~ a>_{lex} b$,
\end{tabbing}
where the lexicographical ordering $>_{lex}$ is defined
 by   $a>_{lex}b$ iff the leftmost nonzero $a_k-b_k>0$.
Other orderings  can be used in the SF package.

For the reduction of Feynman diagrams to the  minimal set 
of basic integrals we found that the following ordering 
is more  efficient than Tresse ordering:
\begin{tabbing}
\quad\=     \quad\= (i) \quad\=  $p>q$ \\
or \> $~~~$ \> (ii) \>  $ $p=q$ $~~ {\rm and}~~  $ord(a)>ord(b),$   \\
or \> $~~~$  \> (iii) $ord(a)=ord(b),~~ $p=q$ ~~{\rm and}~~ a>_{lex} b$.\\
\end{tabbing}

This ordering allows one to exclude all functions with
shifts in $d$ from the minimal set of differential equations.

Input for the SF algorithm may be any linear system of PDEs
for the vector function $V=\{ V^1,...,V^p \}$ depending
on $m$ independent variables $x_1, ..., x_m$.
In our case $V^q$ will be a scalar integral, the same integral
with different shifts in $d$ and integrals obtained from the given
integral by contracting lines. As was mentioned in the previous
section integrals with contracted lines can be treated
as classification functions. As independent variables
$x_1, ..., x_m$ one can take  masses or their ratios.
We assume that all lines of the diagram have different masses.

The main result of the application of the SF algorithm will be a
set of leading derivatives
\begin{equation}
D_bV^p=f^p_b,
\label{leader}
\end{equation}
where $f_b^p$ are explicitly given functions of $x_i$ and
derivatives of $V^q$ such that

i) the derivative on the l.h.s. of (\ref{leader})
 is strictly higher in ordering $>_s$ than any derivative on the 
r.h.s. of (\ref{leader})

ii) no derivative appears on both the
l.h.s. and r.h.s.  of (\ref{leader})

iii) the derivatives on the l.h.s. and r.h.s.
  of (\ref{leader}) are all distinct

iv) no derivative in (\ref{leader}) is a nontrivial
derivative of any derivative in the l.h.s. of
(\ref{leader}).

v) the integrability conditions of (\ref{leader}) are identically
satisfied modulo all lexicographic substitutions which follow from
(\ref{leader}).

To clarify  the last statement we remind the reader
of the definition of integrability (consistency)
conditions. For any distinct pair of equations
\begin{equation}
D_aV^p=f^p_a,~~~~D_bV^p=f^p_b,
\end{equation}
and for any $c=(c_k) \in N^m$, $c_k \geq \max \{a_k,b_k\}$
the consistency conditions are:
\begin{equation}
D_cV^p-D_cV^p=D_{c-a}f_a^p-D_{c-b}f^p_b=0.
\end{equation}
Derivatives on the r.h.s. of (\ref{leader}) are called 
{\it parametric} derivatives i.e. by definition a 
parametric derivative $D_hV^p$ is a derivative that cannot be
obtained by differentiation of any leading derivative.  

The standard form of a linear system of PDEs is achieved by repeating
the following process until all conditions of the standard form
are satisfied:

i) isolate, and solve for, the highest order
(leading) derivatives of each equation

ii) substitute back from (i) throughout the rest of the
 system

iii) append new equations resulting from the integrability
conditions.

The output of the Standard Form algorithm (\ref{leader})
can be identified with our recurrence relations as follows:
\begin{itemize}
\item the set of equations for leading derivatives
 corresponds  to a minimal set of recurrence relations
\item parametric derivatives are our basic integrals
\item  the dimension of the solution
space of a system is equal to the number of
different parametric derivatives in the r.h.s. of
(\ref{leader}).
\end{itemize}

To prove finiteness of the dimension of the solution space
we refer to the known fact that Feynman integrals are holonomic
functions.
Higher order derivatives (integrals corresponding
to scalar diagrams with dots on lines and/or shifts in $d$)
can be obtained by differentiating the set of
leading derivatives (\ref{leader}).
In the Standard Form package \cite{Reid3}) there is a special
subroutine for reducing higher derivatives to parametric ones.
Thus, in principle, after tensor reduction  this subroutine
can be directly applied to the evaluation of Feynman diagrams.

Using the Standard Form package we performed
classification of the generalized recurrence
relations for the one-loop propagator type integrals
and for the simplest two-loop vertex integral with four  lines.

Unfortunately the Standard Form package \cite{Reid3}
(written in {\tt Maple})
is not powerful enough and cannot be used for the calculation
of real diagrams in gauge theories.

We believe that the implementation of the Standard Form algorithm
in other computer languages will give better performance
and can be used in practical calculations.

In the present paper we gave only a short description
of the general scheme for evaluating arbitrary
Feynman diagrams. Details and developments of the
algorithms presented will be published elsewhere.

\section{\em Acknowledgments. }

I am grateful to  V. Gerdt for many useful discussions and
advices. I am also grateful to J. Fleischer and F. Jegerlehner 
for numerous discussions and suggestions. I am thankful to V. Gerdt
and J. Fleischer for carefully reading the manuscript. 
I would like to thank the organizers of the Rheinsberg
conference T.~Riemann and J.~Bl\"umlein. 
Support from BMBF project PH/05-7BI92P 9 is gratefully acknowledged.

\end{document}